\documentclass[aps,prd,preprint,a4paper,showpacs,showkeys,superscriptaddress]{revtex4-1}

\usepackage{latexsym}
\usepackage{amsmath,amssymb}
\usepackage{graphicx}
\usepackage{subfigure}
\usepackage{xcolor}

\usepackage{acronym}

\usepackage{hyperref}     
\hypersetup{colorlinks,%
  citecolor=blue,%
  linkcolor=cyan,%
  pdftex}

\acrodef{AMPS}[AMPS]{Almheiri, Marolf, Polchinski, and Sully}
\acrodef{CGHS}[CGHS]{Callan-Giddings-Harvey-Strominger}
\acrodef{RST}[RST]{Russo-Susskind-Thorlacius}
\acrodef{GEMS}[GEMS]{global embedding in Minkowski spacetime}
\acrodef{AdS}[AdS]{anti-de Sitter}
\acrodef{SAdS}[SAdS]{Schwarzschild anti-de Sitter}

\begin{document}


\title{Test of quantum atmosphere in the dimensionally reduced Schwarzschild black hole}

\author{Myungseok Eune}%
\email[]{eunems@smu.ac.kr}%
\affiliation{Gyedang College of General Education, Sangmyung
  University, Cheonan, 31066, Republic of Korea}%

\author{Wontae Kim}%
\email[]{wtkim@sogang.ac.kr}%
\affiliation{Department of Physics, Sogang University, Seoul, 04107,
  Republic of Korea}%

\date{\today}

\begin{abstract}
  It has been suggested by Giddings that the origin of Hawking
  radiation in black holes is a quantum atmosphere of near-horizon
  quantum region by investigating both the total emission rate and the
  stress tensor of Hawking radiation.  Revisiting this issue in the
  exactly soluble model of a dimensionally reduced Schwarzschild black
  hole, we shall confirm that the dominant Hawking radiation in the
  Unruh vacuum indeed occurs at the quantum atmosphere, not just at
  the horizon by exactly calculating the out-temperature responsible
  for outgoing Hawking particle excitations.  Consequently we show
  that the out-temperature vanishes at the horizon and has
  a peak at a scale whose radial extent is set by the horizon radius,
  and then decreases to the Hawking temperature at infinity.  We also
  discuss bounds of location of the peak for the out-temperature in our
  model.
\end{abstract}



\maketitle


\section{Introduction}
\label{sec:intro}

Hawking radiation as information carrier \cite{Hawking:1974sw} leads
to black hole complementarity in such a way that there would be no
contradictory physical observations between static and freely falling
observers \cite{Susskind:1993if}.  In connection with black hole
complementarity, one of the solutions to the firewall
paradox~\cite{Almheiri:2012rt} is that the infalling observer crossing
the horizon could find the firewall of high frequency quanta after the
Page time \cite{Page:1993wv}.  Thus, the Hawking radiation at the
horizon should be highly excited beyond the Planckian scale.  The
existence of the firewall might also be explained by the Tolman
temperature \cite{Tolman:1930zza,Tolman:1930ona} since the Hawking
radiation at infinity is ascribed to the infinitely blue-shifted
radiation at the horizon.

On the other hand, Unruh showed numerically that the process
of thermal particle creation is low-energy behavior so that the
highest frequency mode does not matter for the thermal emission
\cite{Unruh:1994zw}.  It was also claimed that the Hawking radiation
can be retrieved by an alternative Boulware accretion scenario without
recourse to a pair creation scenario at the horizon
\cite{Israel:2015ava}.
Recently, Giddings raised a refined question regarding the origin of
the Hawking radiation in the Unruh vacuum \cite{Giddings:2015uzr}.  He
investigated both the total emission rate and the stress tensor of
Hawking radiation and then concluded that the origin of Hawking
radiation is the near-horizon quantum region of the quantum atmosphere
whose radial extent is set by the horizon radius scale.  Subsequently,
there have been some related works to the quantum atmosphere; analyses
of the stress tensor and the effective temperature
\cite{Kim:2016iyf,Dey:2017yez,Dey:2019ugf}, and calculations of
emission rate of Hawking radiation in arbitrary dimensions
\cite{Hod:2016hdd}.

In particular, from the analysis of a stress tensor,
it was claimed that in the Unruh vacuum
the Tolman temperature near the horizon does not originate from the out-going particles
but the in-going particles from the fact that the negative influx would transition
to the positive outward flux over the quantum region outside the horizon  \cite{Giddings:2015uzr}.
It means that the out-going Hawking radiation originates from the quantum atmosphere. This claim was also discussed by employing the local temperature responsible for the out-going particles \cite{Kim:2016iyf};
thus the local temperature related to the out-going particles must be finite over
the whole region, in particular, it has a peak at a macroscopic distance outside the horizon.
The crucial difference from conventional results comes from a modification of the Stefan-Boltzmann law
for the out-going particles. In Ref.~\cite{Dey:2017yez}, the authors also advocated
the quantum atmosphere with two different arguments. Heuristically,
the first was based on the gravitational Schwinger effect for particle production by the tidal force outside a black hole horizon.
Next, the second argument of our concern made use of a calculation of the stress tensor
to derive the energy density for an observer at a constant Kruskal position in order to investigate the quantum atmosphere. However, the second argument relied on the observer in Kruskal coordinates,
which are not free-fall coordinates at a finite  distance from the horizon.
Subsequently, this issue was resolved by using a free-fall coordinate
system without acceleration \cite{Dey:2019ugf}.
Furthermore, the adiabaticity of the test field modes, which allows us to test where the Wentzel-Kramers-Brillouin
breaks down, was taken as an indicator of a particle creation process.
The authors found a peak in the violation condition for the field modes; then the condition agrees to
the peak of the discrepancy between the energy density and the effective temperature.
This discrepancy was regarded as a signal for the location of quantum atmosphere
where a particle creation process is taking place.

In this work we would like to revisit the spatial origin of Hawking
radiation by investigating the local temperature responsible for the outgoing Hawking
particles in a dimensionally reduced Schwarzschild black hole which is
more or less realistic and exactly soluble model.  The organization of
this paper is as follows.  In Sec.~\ref{sec:s-wave} we study expectation values of the
stress tensor for scalar fields from the one-loop
effective action in the dimensionally reduced model
based on Refs.~\cite{Nojiri:1997sr,Kummer:1997jr,Kummer:1998sp,Dowker:1998bm}.
Then in Sec.~\ref{Three vacuum states} the stress tensor in
thermodynamic equilibrium states proposed by Balbinot and
Fabbri~\cite{Balbinot:2003fv} will be extended to the stress tensor in
the Unruh vacuum of nonequilibrium state.  In
Sec.~\ref{sec:perfect.fluid:2d} we first calculate the local
temperature in the Hartle-Hawking vacuum and then find it is finite everywhere. Next we show that the local temperature in
the Unruh vacuum is just the Tolman temperature, but it consists of
the influx and outward flux. Since the only outgoing modes contribute
to the Hawking temperature at spatial infinity,
we should discard the influx from the Tolman temperature and
then justify the out-temperature purely characterized by the outgoing Hawking
radiation.  Eventually the out-temperature turns out to be the same as the local
temperature in the Hartle-Hawking vacuum; it vanishes at the horizon
and has a peak at a scale whose radial extent is set by the horizon
radius in the quantum atmosphere, and then approaches the Hawking
temperature at infinity. Furthermore we find the lower bound and the
upper bound of the peak.  Finally conclusion will be given in
Sec.~\ref{sec:discussion}.

\section{Dimensional reduction of the Einstein-Hilbert action}
\label{sec:s-wave}

Let us start with the Einstein-Hilbert action and the matter action
defined by
\begin{align}
  I_{\rm EH} &= \frac{1}{16\pi G} \int d^4 x \sqrt{-g^{\mathrm{(4D)}}}
               R^{\mathrm{(4D)}}, \label{eq:action:EH:4d}  \\
  I_{\rm m}^{\rm (4D)} &= - \frac{1}{8\pi} \int d^4 x \sqrt{-g^{\mathrm{(4D)}}}
                         \sum_{i=1}^N \left( \nabla f_i \right)^2, \label{eq:action:matter:4d}
\end{align}
where $f_i$ is a scalar field and $N$ is the number of the scalar fields.
In the spherically symmetric space, the four-dimensional line element
can be written as
$ ds^2_{\mathrm{(4D)}} = ds^2 + (1/\lambda^2) e^{-2\phi}
d\Omega_{2}^2$ with $ds^2 = g_{\mu \nu} (x) dx^\mu dx^\nu$ and
$\phi = \phi(x)$, where $\lambda^2 = \pi/(2G)$, $e^{-\phi}$ is the
radius, and $d\Omega_{2}$ is the two-dimensional solid angle. Then,
the four-dimensional actions \eqref{eq:action:EH:4d} and
\eqref{eq:action:matter:4d} reduce to
\begin{align}
  I &= \frac{1}{4G} \int d^2 x \sqrt{-g} e^{-2\phi} \left[
      R + 2 (\nabla\phi)^2 + 2\lambda^2 e^{2\phi}
      \right], \label{eq:action:0:2d} \\
  I_{\rm m} &= - \frac12 \int d^2 x \sqrt{-g} e^{-2\phi} \sum_{i=1}^N
              \left(\nabla f_i \right)^2. \label{eq:action:matter:2d}
\end{align}
Solving the equations of motion from the actions
\eqref{eq:action:0:2d} and \eqref{eq:action:matter:2d}, one can obtain
the two-dimensional vacuum solution for a static black hole described
by
\begin{align}
  ds^2 &= - f(r) dt^2 + \frac{dr^2}{f(r)}, \label{ds:metric:2d}
\end{align}
where $f(r) = 1 - (2GM)/r$, $\phi = -\ln (\lambda r)$, and $f_i =0$.
The event horizon is located at $r_h = 2GM$.  In terms of the
light-cone coordinates, the line element~\eqref{ds:metric:2d} is also
written as
\begin{align}
  ds^2 = - e^{2\rho} d\sigma^+ d\sigma^-, \label{ds:tortoise}
\end{align}
where $\rho = (1/2) \ln f$ and $\sigma^\pm = t \pm r^*$.  The tortoise
coordinate is defined by $r^* = r + r_h \ln \left( rf/r_h \right)$, and its
inverse relation is given by $r = r_h [1+ W(e^{r^*/r_h-1})]$, where
the function $W(z)$ is the Lambert $W$ function satisfying
$z = W e^{W}$.

On the other hand, the one-loop effective action for the scalar fields
in Eq.~\eqref{eq:action:matter:2d} is obtained as
\begin{align}
  \bar{\Gamma} &= - \frac{N}{96\pi} \int d^2 x \sqrt{-g} \left[ R
                 \frac{1}{\Box} R - 12 R \frac{1}{\Box} (\nabla \phi)^2 + 12
                 \phi R \right], \label{eq:action:1:nonlocal}
\end{align}
and the trace anomaly of the
semi-classically quantized stress tensor
reads as \cite{Nojiri:1997sr,Dowker:1998bm,Kummer:1997jr,Kummer:1998sp}
\begin{align}
  \langle T_\mu^\mu \rangle &= \frac{N}{24\pi} \left[ R - 6 (\nabla
                              \phi)^2 + 6 \Box \phi \right].\label{trace.EM:cov}
\end{align}
However, the flux in the Hartle-Hawking vacuum is unfortunately
negative at spatial infinity. In order to evade the negative flux at
infinity \cite{Mukhanov:1994ax,Kummer:1998dc}, Balbinot and Fabbri \cite{Balbinot:2003fv} considered the Weyl
invariant action not affecting the trace anomaly such as
\begin{align}
  \frac{N}{96\pi} \left(b^2 - 36\right)\int d^2 x \sqrt{-g} (\nabla \phi)^2
  \frac{1}{\Box} (\nabla \phi)^2 \label{action:conf.inv}
\end{align}
which is arbitrary but phenomenologically sensible in that Eq.
\eqref{action:conf.inv}
will provide the desired expression of the Hawking radiation at infinity.
The constant $b$ was determined as $b=2\sqrt3$, which
can be identified with $b=4 \sqrt{3\pi}\ell_1$ in
Ref.~\cite{Balbinot:2003fv}. Combining
Eq.~\eqref{eq:action:1:nonlocal} with Eq.~\eqref{action:conf.inv}, we
will take the one-loop effective action as
\begin{align}
  \Gamma &= - \frac{N}{48\pi} \int d^2 x \sqrt{-g} \left[
           \psi \left( R - 6(\nabla\phi)^2 \right) + \frac12 (\nabla\psi)^2 +
           6 \phi R - \frac12 (\nabla \chi)^2 - 2 \sqrt 3 \chi
           (\nabla \phi)^2 \right], \label{action:1:new}
\end{align}
where the two auxiliary scalar fields $\psi$ and $\chi$ satisfy
\begin{align}
  \Box \psi = R - 6 (\nabla \phi)^2,\quad
  \Box \chi = 2 \sqrt 3 (\nabla \phi)^2. \label{eom:chi:new}
\end{align}

From the localized action \eqref{action:1:new}, the quantum-mechanical
stress tensor is easily obtained as
\begin{align}
  \langle T_{\mu\nu} \rangle
  &= \frac{N}{24\pi} \bigg[ -\nabla_\mu \nabla_\nu \psi + \frac12 \nabla_\mu
    \psi \nabla_\nu \psi - 6 \psi \nabla_\mu \phi \nabla_\nu \phi - 6
    \left(\nabla_\mu \nabla_\nu \phi -   g_{\mu\nu} \Box \phi
    \right)-\frac12 \nabla_\mu \chi \nabla_\nu \chi  \notag \\
  &\quad  - 2\sqrt3 \chi \nabla_\mu \phi \nabla_\nu \phi+
    g_{\mu\nu} \left( \Box \psi - \frac14 (\nabla \psi)^2 + 3\psi
    (\nabla\phi)^2   +  \frac14 (\nabla \chi)^2 + \sqrt 3 \chi (\nabla
    \phi)^2 \right)\bigg].\label{T.ab}
\end{align}
In the conformal gauge \eqref{ds:tortoise}, Eq.~\eqref{eom:chi:new} is
written as
\begin{align}
  \partial_+ \partial_- \psi
  = - 2 \partial_+ \partial_- \rho - 6 \partial_+ \phi \partial_-
  \phi, \quad  
  \partial_+ \partial_- \chi
  = 2 \sqrt 3 \partial_+\phi \partial_- \phi, \label{eeom:chi:conformal}
\end{align}
and the general solutions for $\psi$ and $\chi$ are easily obtained as
\begin{align}
  \psi &= -2 \rho + 6 \xi + \omega_+(\sigma^+) + \omega_- (\sigma^-), \label{sol:psi:new}\\
  \chi &= -2 \sqrt 3 \xi + \eta_+ (\sigma^+) + \eta_- (\sigma^-), \label{sol:chi:new}
\end{align}
where $\xi = \ln \left((r / r_h) \sqrt{f} \right)$ and
$\omega_\pm (\sigma^\pm)$ and $\eta_\pm(\sigma^\pm)$ are arbitrary
holomorphic/antiholomorphic functions determined by some boundary
conditions of vacuum states. Plugging Eqs.~\eqref{sol:psi:new} and
\eqref{sol:chi:new} into Eq.~\eqref{T.ab}, the components of the
stress tensor \eqref{T.ab} are expressed as
\begin{align}
  \langle T_{+-} \rangle &= \frac{N}{12\pi} \left[-\partial_+ \partial_-\rho -
                           3 \partial_+ \phi \partial_- \phi + 3
                           \partial_+ \partial_- \phi \right], \label{T.+-:new}\\
  \langle T_{\pm\pm} \rangle &= \frac{N}{12\pi} \bigg[\partial_\pm^2
                               \rho - (\partial_\pm \rho)^2 + 6\rho
                               (\partial_\pm \phi)^2 - 3
                               (\partial_\pm^2 \phi - 2 \partial_\pm
                               \rho \partial_\pm \phi) -3
                               \partial_\pm^2 \xi \notag \\
                         &\quad + 6 (\partial_\pm  \xi)^2 - 288 \xi
                           (\partial_\pm \phi)^2 +
                           \partial_\pm \xi (3\partial_\pm \omega_\pm
                           + \sqrt 3 \partial_\pm \eta_\pm ) \notag \\
                         &\quad - (\partial_\pm \phi)^2
                           \left( (3\omega_+ + \sqrt 3 \eta_+ )+(3\omega_- +\sqrt 3 \eta_-) \right)  - t_\pm \bigg], \label{T.++:new}
\end{align}
where we defined
$t_\pm = (1/2) \partial_\pm^2 \omega_\pm - (1/4) (\partial_\pm
\omega_\pm)^2 + (1/4) (\partial_\pm \eta_\pm)^2$.  Here, in order for
the stress tensor~\eqref{T.++:new} to be static, one of the simplest
solutions in Eqs.~\eqref{sol:psi:new}
and~\eqref{sol:chi:new} might be chosen as $\omega_\pm = \pm (1/2) c_1 \sigma^\pm + d_1$
and $\eta_\pm = \pm (1/2) c_2 \sigma^\pm + d_2$~\cite{Balbinot:2003fv},
where $c_1$, $d_1$, $c_2$, $d_2$ are constants. The two constants of
$d_1$ and $d_2$ can be written as one constant without affecting the
stress tensor due to the translation symmetry in it; however, this
choice of $\omega_\pm$ and $\eta_\pm$ always results in $t_+ = t_-$ in
Eq.~\eqref{T.++:new}, which describes only equilibrium states such as
the Hartle-Hawking and the Boulware vacuum. In order to incorporate
non-equilibrium vacuum, we impose a less strong condition for the
auxiliary fields as
\begin{equation}
  3\omega_\pm + \sqrt 3 \eta_\pm = \pm  \frac{C}{2} \sigma^\pm -
  \frac{1}{2}D, \label{less}
\end{equation}
which also successfully makes Eqs.~\eqref{T.+-:new} and
\eqref{T.++:new} static, where $C$ and $D$ are constants.
Note that arbitrary holomorphic/antiholomorphic
integration functions  in
Eqs. \eqref{sol:psi:new} and \eqref{sol:chi:new}
can be written as $\omega_\pm =\sum_{n=0}^{n=\infty} a_n^{\pm} (\sigma^\pm)^n$ and $\eta_\pm=\sum_{n=0}^{n=\infty}b_n^{\pm}
(\sigma^\pm)^n$ with the constants $a_n$ and $b_n$,
and thus Eq. \eqref{less} can be obtained by choosing $ 3a_0^\pm +\sqrt{3}b_0^\pm =-(1/2)D$ and
$3a_1^\pm +\sqrt{3}b_1^\pm =\pm C/2  $.
Eventually, the integration constant $D$ comes from
the zero modes of $\omega_\pm$ and $\eta_\pm$.

Using
Eqs.~\eqref{ds:tortoise} and \eqref{less}, one can write
Eqs.~\eqref{T.+-:new} and \eqref{T.++:new} as
\begin{align}
  \langle T_{+-} \rangle
  &= \frac{Nf}{48\pi} \left[\frac12 f'' + 3 \frac{f'}{r}
    \right], \label{T:+-:new:f} \\
  \langle T_{\pm\pm} \rangle
  &= \frac{N}{48\pi} \bigg[ \frac12 ff'' - \frac14 (f')^2 + \frac{3}{2r^2}
    (1 + f)^2 + \frac{C}{2r}(1+f) - 4 t_\pm \notag \\
  &\quad + \frac{f^2}{2r^2} \left( - (6+ 2C r_h)
    \ln f  - (24+ 2Cr_h) \ln \frac{r}{r_h} - 2Cr +2 D \right)
    \bigg], \label{T:++:new:f}
\end{align}
where the integration function is written as
$t_\pm = (1/2) \partial_\pm^2 \omega_\pm + (1/2) (\partial_\pm
\omega_\pm)^2 \mp (\sqrt 3 C/4) \partial_\pm \omega_\pm + C^2/48$.

\section{Three vacuum states}
\label{Three vacuum states}

We consider three vacuum states; the Boulware~\cite{Boulware:1974dm},
the Hartle-Hawking~\cite{Israel:1976ur,Hartle:1976tp}, and the Unruh
vacuum~\cite{Unruh:1976db}; thus $C=\{ C_{\textrm B}, C_{\rm HH}, C_{\rm U} \}$, $D=\{ D_{\textrm B}, D_{\rm HH}, D_{\textrm U} \}$, and $t_{\pm}^{\rm B,\rm HH, \rm U}$ will be determined for a given
vacuum state.  Firstly in the Boulware vacuum there
are no influx and outward flux at spatial infinity, so that we can
choose $t_\pm^{\rm B} = 0$ in Eq.~\eqref{T:++:new:f}.  Furthermore in
the limit of the Minkowski spacetime of $M=0$ \cite{Balbinot:2003fv},
Eq.~\eqref{T:++:new:f} should vanish as
\begin{align}
  \langle T_{\pm\pm} \rangle_\mathrm{B}
  &\to \lim_{r_h \to 0} \frac{N}{48\pi} \left[\frac{12+3
    C_\mathrm{B}r_h  +2 D_{\rm B}}{2r^2}
    - \frac{12 + C_{\rm B}r_h}{r^2} \ln \frac{r}{r_h} \right] = 0,
    \label{T:++:B:Min}
\end{align}
which determines $C_{\rm B} = -12 /r_h$ and
$D_{\rm B}=12$.
Consequently, we get
\begin{align}
  \langle T_{\pm\pm} \rangle_\mathrm{B}
  &= \frac{N x^2}{192\pi r_h^2} \left[ 19x(3x-4)+36  f^2
    \ln f \right], \label{T:++:new:B:final}
\end{align}
where $f=1-x$ with $x = r_h/r$. All constants can be completely
fixed in the Boulware vacuum thanks to the additional condition
\eqref{T:++:B:Min}.

Secondly in the Hartle-Hawking vacuum both the influx and outward
flux are finite at the horizon, in other words,
$\langle T^{\pm\pm} \rangle_{\mathrm{HH}}$ is regular. It requires
that $\langle T_{\pm\pm} \rangle_\mathrm{HH}$ should vanish
necessarily at the horizon since
$\langle T^{\pm\pm} \rangle_{\mathrm{HH}} = \langle T_{\pm\pm}
\rangle_\mathrm{HH} / f^2$, and thus we can fix the boundary
conditions as $t_\pm^{\rm HH} = -1/(4r_h)^2$ and $C_{\rm HH}=-3/r_h$.
Plugging these conditions into Eq.~\eqref{T:++:new:f}, we can obtain
\begin{align}
  \langle T_{\pm\pm} \rangle_{\mathrm{HH}}
  &= \frac{Nf^2}{192\pi r_h^2} \left[1 + 2x + x^2(9 +4 D_{\rm HH} + 36 \ln x)
    \right],\label{T:++:new:HH:final}
\end{align}
where  $D_{\mathrm{HH}}$ is not fixed
by the Hartle-Hawking boundary condition. Note that one cannot
require the condition like Eq.~\eqref{T:++:B:Min} in the
Hartle-Hawking vacuum since there does not exist the limit of the
Minkowski spacetime of $M=0$ unlike the case of the Boulware vacuum. Note that
Eqs.~\eqref{T:++:new:B:final} and \eqref{T:++:new:HH:final} are
exactly the same as the results in Ref.~\cite{Balbinot:2003fv}.
As expected, the flux at infinity is well-defined as
$\langle T_{\pm\pm} \rangle_{\mathrm{HH}} \to N( \pi/12)
T_\mathrm{H}^2$, where $T_\mathrm{H}=1/(4 \pi r_h)$.

Finally in the Unruh vacuum there is no influx at spatial
infinity and the outward flux is finite at horizon, specifically,
$\langle T^{--} \rangle_\mathrm{U}$ vanishes at infinity and
$\langle T^{++} \rangle_{\mathrm{U}}$ is regular on the horizon, so
that the boundary conditions are chosen as $t_+^{\rm U} = 0$ and
$t_-^{\rm U} = -(4r_h)^{-2}$ with $C_{\rm U}=-3/r_h$.  Then the
resulting stress tensor can be newly obtained as
\begin{align}
  \langle T_{++} \rangle_\mathrm{U}
  &= \frac{N x^2}{192\pi r_h^2} \left[ 6 +4 D_{\rm U} - 8(2 + D_{\rm U}) x
    +(9 +4 D_{\rm U}) x^2 +36 f^2 \ln x \right], \label{T:++:new:U:final} \\
  \langle T_{--} \rangle_\mathrm{U}
  &= \frac{Nf^2}{192\pi r_h^2} \left[1 + 2x + x^2(9 +4 D_{\rm U} + 36 \ln
    x) \right]. \label{T:--:new:U:final}
\end{align}
Near the horizon the influx is negative finite as
$\langle T_{++} \rangle_\mathrm{U} \to -N/(192\pi r_h^2)$, while $\langle T_{--} \rangle_\mathrm{U} \to 0$.  At spatial infinity
$\langle T_{++} \rangle_\mathrm{U} \to 0$, while
$\langle T_{--} \rangle_\mathrm{U} \to N( \pi/12) T_\mathrm{H}^2$.

\section{Local temperatures in equilibrium and non-equilibrium state}
\label{sec:perfect.fluid:2d}

In this section, we calculate the local temperature in order to study
the origin of Hawking radiation in the evaporating black hole.
For this purpose, the
proper velocity of radiation flow can be found by solving the geodesic
equation of $u^\alpha \nabla_\alpha u^\mu = 0$, where the proper
velocity $u^\mu$ is defined by $u^\mu = dx^\mu/d\tau$ with a proper
time $\tau$.  Then the proper velocity can be obtained as
\begin{align}
  u^\mu = \left( \frac{\sqrt{f(r_0)}}{f(r)}, \pm \sqrt{f(r_0) - f(r)}
  \right). \label{u:sol:ff.rest:general}
\end{align}
If the frame is released from rest at an arbitrary point of $r=r_0$,
then Eq.~\eqref{u:sol:ff.rest:general} reduces to
$u^\mu = \left( 1/\sqrt{f}, 0 \right)$.  Next, we define the local
quantities related to the stress tensor as
\begin{align}
  \varepsilon = \langle T_{\mu\nu} \rangle u^\mu u^\nu, \quad  p =
  \langle T_{\mu\nu} \rangle  n^\mu n^\nu,
  \quad \mathcal{F}= - \langle T_{\mu\nu} \rangle  u^\mu n^\nu, \label{proper:T.ab}
\end{align}
where $\varepsilon$, $p$, and $\mathcal{F}$ are the proper energy
density, pressure, and flux, respectively. Note that $n^\mu$ is a
spacelike unit vector normal to $u^\mu$ given by
$n^\mu = (0, \sqrt{f} )$.  In the light-cone coordinates, the energy
density and flux in Eq.~\eqref{proper:T.ab} can also be expressed as
\begin{align}
  \varepsilon &= \frac{1}{f} (\langle T_{++}  \rangle + \langle T_{--}
                \rangle + 2 \langle T_{+-} \rangle
                ), \label{rho:tortoise}\\
  \mathcal{F} &= - \frac{1}{f} (\langle T_{++}  \rangle - \langle
                T_{--} \rangle ), \label{flux:tortoise}
\end{align}
where we used $u^\pm = 1/\sqrt{f}$ and $n^\pm = \pm 1/\sqrt{f}$ in the
light-cone coordinates.

Before studying the local temperature in the Unruh vacuum,
we first investigate the Stefan-Boltzmann law in the Hartle-Hawking
vacuum in order to relate the proper energy density
\eqref{rho:tortoise} to the local temperature along the line of
Refs.~\cite{Tolman:1930zza,Tolman:1930ona}; however, we will take into
account the trace anomaly in deriving the Stefan-Boltzmann law as compared to the conventional procedure.
Let us start with the first law of thermodynamics written as
$dU = TdS - pdV $ where $U$, $S$, $V$, and $T$, and $p$ are the
internal energy, entropy, volume, temperature, and pressure of a
system. At a fixed temperature, the first law of thermodynamics can be
rewritten as
\begin{align}
  \left(\frac{\partial U}{\partial V} \right)_T = T
  \left(\frac{\partial S}{\partial V} \right)_T - p, \label{1st.law:rewriting}
\end{align}
where the left-hand side is the energy density, \textit{i.e.,}
$\varepsilon=\left(\partial U /\partial V \right)_T$. Using the
Maxwell relation of
$\left(\partial S/\partial V \right)_T = \left( \partial p/\partial T
\right)_V$, one can see that Eq.~\eqref{1st.law:rewriting} becomes
\begin{align}
  \varepsilon = T \left(\frac{\partial p}{\partial T} \right)_V - p. \label{1st.law:diff.eq}
\end{align}
Assuming the trace of the stress tensor does not vanish as seen from
Eq.~\eqref{trace.EM:cov}, we can write
\begin{align}
  \langle T^\mu_\mu  \rangle = -\varepsilon + p, \label{trace:fluid}
\end{align}
by using Eq.~\eqref{proper:T.ab}.  From
Eqs.~\eqref{1st.law:diff.eq} and ~\eqref{trace:fluid}, one can
eliminate the pressure term and thus obtain the first order
differential equation as
\begin{align}
  \label{key}
  T \left(\frac{\partial \varepsilon}{\partial T}\right)_V
  -2\varepsilon = \langle T^\mu_\mu \rangle ,
\end{align}
where we used the fact that the trace anomaly is independent of
temperature~\cite{BoschiFilho:1991xz}.  Then, the energy density is
solved as \cite{Gim:2015era}
\begin{align}
  \varepsilon = \gamma T^2 - \frac12 \langle T^\mu_\mu
  \rangle, \label{energy:thermo}
\end{align}
where $\gamma$ is an integration constant which turns out to be the
Stefan-Boltzmann constant. Note that the Stefan-Boltzmann law
\eqref{energy:thermo} modified by the trace anomaly naturally reduces
to the well-known Stefan-Boltzmann law in
Refs.~\cite{Tolman:1930zza,Tolman:1930ona} if
$\langle T^\mu_\mu \rangle =0$.  The second term in
Eq.~\eqref{energy:thermo} will modify the behavior of the local
temperature significantly near the horizon.
\begin{figure}[pt]
  \begin{center}
     \includegraphics[width=0.8\textwidth]{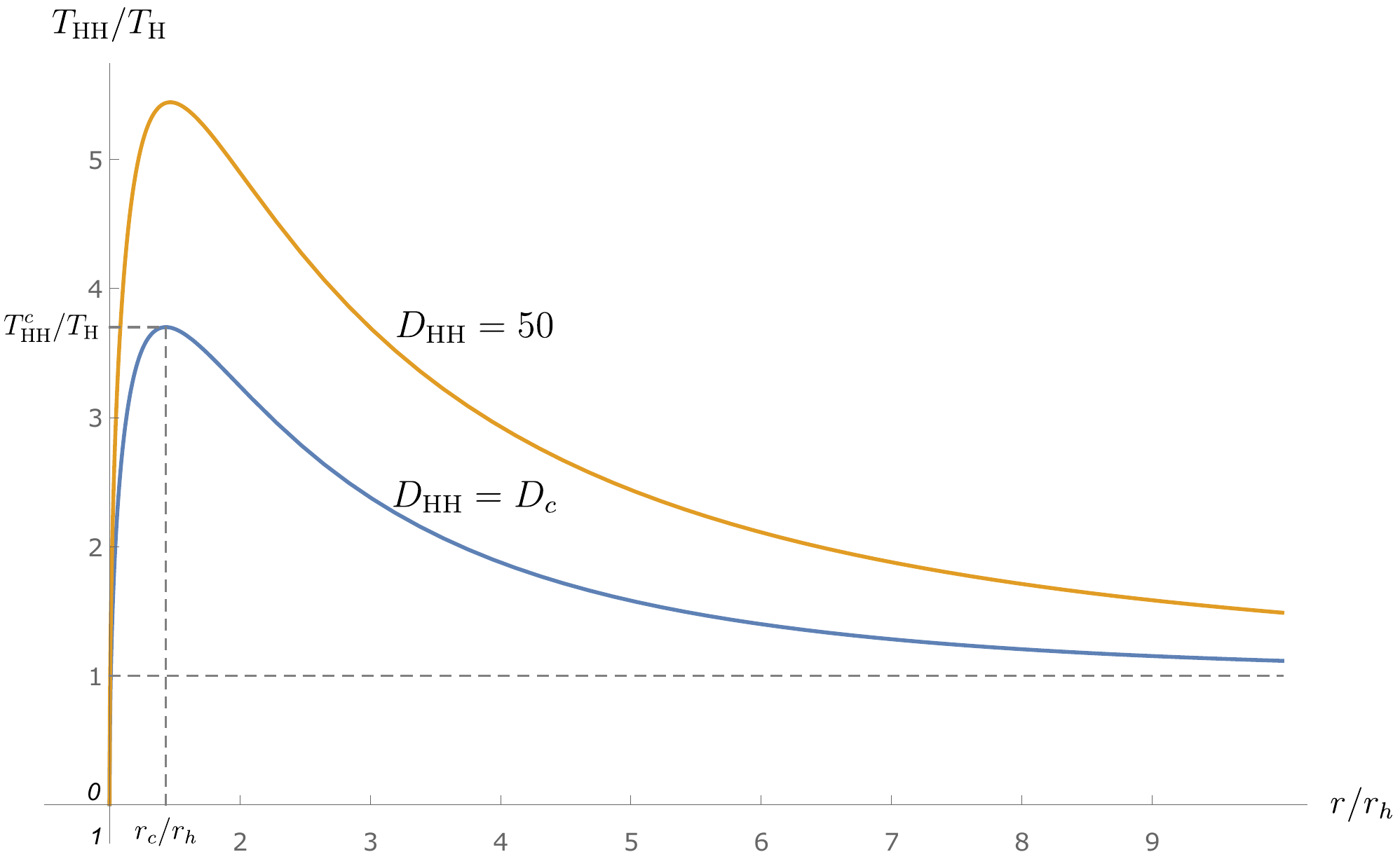}
  \end{center}
  \caption{The local temperatures $T_{\rm HH}$ in the Hartle-Hawking vacuum are
    plotted for $D_{\rm HH} \ge D_c\approx 23.03$, where the
    temperatures are real and monotonically decreasing after each
    peaks.  In case of $D_{\rm HH} = D_c$, the ratio of the
    maximum local temperature to the Hawking temperature is
    $T_{\rm HH}^c/T_\mathrm{H} \approx 3.70 $ and it occurs at
    $r_c /r_h\approx 1.43$. The location of the peak temperature is
    shifted to the right from $r_c$ as
    $D_\mathrm{HH}$ is getting large.}
  \label{fig:HH:temperature}
\end{figure}

Plugging Eq.~\eqref{rho:tortoise} into Eq.~\eqref{energy:thermo}, we
obtain the local temperature in the Hartle-Hawking vacuum as
\begin{align}
  \gamma T^2_{\rm HH} = \frac{1}{f} (\langle T_{++}
  \rangle_{\mathrm{HH}} + \langle T_{--} \rangle_{\mathrm{HH}})
  =\frac{2}{f} \langle T_{\pm\pm}  \rangle_{\mathrm{HH}},  \label{Temp:stress.T}
\end{align}
where
$\langle T_{++} \rangle_{\mathrm{HH}} = \langle T_{--}
\rangle_{\mathrm{HH}}$ from Eq.~\eqref{T:++:new:HH:final}, and
requiring $T_{\rm HH} \to T_{\rm H}$ at infinity determines
$\gamma = \pi N / 6$. The local temperature $T_{\rm HH}$ in the Hartle-Hawking vacuum \eqref{Temp:stress.T} can
be explicitly calculated as
\begin{align}
  T_{\rm HH} (r) &= T_{\mathrm{H}} \sqrt{1- \frac{r_h}{r}} \sqrt{1 +
                   \frac{2r_h}{r} + \left( \frac{r_h}{r} \right)^2 \left(9
                   +4 D_{\rm HH} + 36 \ln \frac{r_h}{r}
                   \right)}, \label{T:HH}
\end{align}
by using Eq.~\eqref{T:++:new:HH:final}.  It is interesting to note
that the local temperature vanishes at the horizon. Of course, it
approaches the expected Hawking temperature at infinity.  The near
horizon limit of the local temperature is very different from that of
the conventional Tolman temperature defined by $T = T_\mathrm{H}/\sqrt{f(r)}$
which is divergent at the horizon.  The zero temperature at the
horizon is compatible with the fact that there exist neither influx
nor outward flux at the horizon in thermal equilibrium.

The constant $D_{\rm HH}$ in Eq.~\eqref{T:HH} is still arbitrary
because it is not enough to fix all constants just by using the
Hartle-Hawking boundary condition imposed on the stress tensor in contrast to the
case of the Boulware vacuum. Thus, it
should be fixed by an additional condition.  If
$D_{\rm HH} < D_0$, the temperature~\eqref{T:HH} is
imaginary in a certain region outside the horizon, where
$D_0 =  -(53 + \sqrt{73})/8 + 9 \ln \left[ (\sqrt{73} -
  1)/2\right] \approx 4.26$.  If
$D_0 \le D_{\rm HH} <D_c$ where
$D_c \approx 23.03$, the temperature is real in the
whole region, but unfortunately it is not decreasing monotonically.
Finally if we take $D_{\rm HH} \ge D_c $, then the
temperature is not only real in the whole region but also
decreasing monotonically as $r$ increases after it reaches a maximum
value at $r_c \approx 1.43 r_h$ (see Fig.~\ref{fig:HH:temperature}).
In this respect we will take $D_{\rm HH} \ge D_c $ from now
on.

Let us now find the locations of peaks from
$\left.\partial_r T_\mathrm{HH} \right|_{r=r_\mathrm{peak}}= 0$ with assuming $D_{\rm HH} \ge D_c $, where
$r_\mathrm{peak}$ is a position at which the maximum temperature
occurs, then we obtain the relation between $D_\mathrm{HH}$ and
$r_\mathrm{peak}$ as
\begin{align}
  D_\mathrm{HH} = \frac{63 - 50 r_\mathrm{peak}/r_h-(
  r_\mathrm{peak}/r_h)^2}{8 ( r_\mathrm{peak}/r_h-3/2)}+18
  \left(\frac{r_\mathrm{peak}}{r_h}-\frac{3}{2}\right) \ln
  \frac{r_\mathrm{peak}}{r_h}. \label{T:HH:peak:alpha.rpeak}
\end{align}
If $r_\mathrm{peak}$ occurs at $r_h$, then $D_\mathrm{HH}=-3$, which is prohibited by the assumption that
$D_{\rm HH} \ge D_c\approx 23.03$.
Hence, for $D_{\rm HH} \ge D_c$, $r_\mathrm{peak}$ should be at a finite distance from the horizon and increases to
$(3/2)r_{h}$ monotonically as $D_{\rm HH}$ increases.
As a result, the peaks of the local temperatures in equilibrium should
lie in
\begin{equation}
  1.43 r_h \lesssim r_\mathrm{peak} < 1.5r_h.
\end{equation}

Now, in the Unruh vacuum, one can calculate the local
temperature by substituting Eqs.~\eqref{T:++:new:U:final} and
\eqref{T:--:new:U:final} into Eq.~\eqref{flux:tortoise} as
\cite{Giddings:2015uzr,Kim:2016iyf,Kim:2017aag}
\begin{align}
  \mathcal{F_\mathrm{U}}  = - \frac{1}{f} \left( \langle T_{++}
  \rangle_\mathrm{U} -  \langle T_{--} \rangle_\mathrm{U} \right)
  = \sigma \left(\frac{T_\mathrm{H}}{\sqrt{f}}\right)^2=\sigma
  T^2_\mathrm{U}, \label{flux:stress}
 \end{align}
 where $\sigma=\gamma/2$.  The local temperature in the Unruh vacuum
 takes exactly the Tolman's form, which is not new; however, it is
 worth noting that the local temperature \eqref{flux:stress} consists
 of the negative influx and the positive outward flux.  Defining
\begin{align}
  \sigma T_{\rm in}^2 =- \frac{1}{f} \langle T_{++} \rangle_\mathrm{U},
  \quad \sigma T_{\rm out}^2= \frac{1}{f}   \langle T_{--}
  \rangle_\mathrm{U}, \label{twofluxes}
\end{align}
we obtain
\begin{equation}
  T_{\rm in}^2 +T_{\rm out}^2 = T^2_\mathrm{U},
\end{equation}
where $T_{\rm in}$ and $T_{\rm out}$ are related to the influx and outward flux, respectively.
Note that the divergence of the Tolman temperature at the horizon
comes from $T_{\rm in}$ since the influx~\eqref{T:++:new:U:final} is
negative finite at the horizon but it is infinitely blue-shifted
there.
Moreover, it turns out that $T_{\rm out}$ is exactly the same as the
local temperature~\eqref{Temp:stress.T} in the Hartle-Hawking vacuum
as seen from Eq.~\eqref{T:++:new:HH:final} and
Eq.~\eqref{T:--:new:U:final} if $D_{\rm U}=D_{\rm HH}$,
which results in
\begin{equation}
  T_{\rm out}=T_{\rm HH}. \label{out}
\end{equation}
Therefore, the local temperature associated with the outgoing Hawking
radiation can be finite everywhere, and its peak occurs at a
macroscopic distance outside the horizon, which means that the main
excitations occur not at the horizon but at the peak in the quantum
atmosphere.

\section{Conclusion}
\label{sec:discussion}

In the dimensionally reduced Schwarzschild black hole, we
found that the divergence of the temperature at the horizon comes from the infinite blue-shift of the negative influx, and
the out-temperature responsible for the Hawking radiation is always
finite, more importantly, its peak occurs in the quantum atmosphere bounded by
$1.43 r_h \lesssim r_\mathrm{peak} < 1.5r_h$, which is indeed the size
of the black hole. It means that the main excitations of Hawking particles
dominantly happens at the
peak but it spreads throughout the whole region well outside the
horizon.  Consequently, in the spherically
reduced Schwarzschild black hole, we confirmed that the origin of the
Hawking radiation is the quantum atmosphere not just at the horizon
from the viewpoint of the local temperature in the semi-classical regime.

As a first comment, we discuss what it happens when the trace anomaly
\eqref{trace.EM:cov} is ignored in our calculations. In
Eq.~\eqref{key}, if we replace $\langle T^\mu_\mu \rangle$ by
$q \langle T^\mu_\mu \rangle$ where $q$ is $0$ or $1$, the temperature is obtained as
$T^2_{\rm HH} = T_\mathrm{H}^2[ f(1 + 2x + x^2(9 + 4 D_{\mathrm{HH}} + 36 \ln
x)) + 8 (1-q) x^3]$.  At the horizon,
$T_{\rm HH} = 2 \sqrt 2 T_\mathrm{H} \sqrt{1-q}$ so that in the
usual case of $q=0$ the temperature does not vanish.
In our case of $q=1$ the temperature vanishes at the horizon,
which means that the modified Stefan-Boltzmann law \eqref{energy:thermo} induced by
the trace anomaly is essential in our atmosphere argument.

Finally, one might wonder what the physical meaning of $D$ is,
in consideration of the fact that three regimes are identified in Sec. \ref{sec:perfect.fluid:2d}
and thus only $D \ge D_c $ seems to be physically relevant. Unfortunately,
we are not well aware of the meaning of the constant $D$ generated from the semiclassical treatment
although $D$ plays the important role in our calculations. Instead,
we notice some intriguing points for further study:
(i) $D$ arises from the solution of the auxiliary scalar fields
to localize the non-local
effective actions~\eqref{eq:action:1:nonlocal} and~\eqref{action:conf.inv}.
What needs to be answered is that it can be a quantum-mechanical hair or not. Otherwise,
we must find a way to fix the constant like the case of Boulware state. (ii) For $D \neq D'$,
the Hartle-Hawking states such as $|\rm HH; D\rangle$
and $|\rm HH; D'\rangle$ are degenerated at the horizon, and thus the 
local temperatures are coincident at the horizon; however,
they depend on $D$ in the bulk region. We hope these issues will be resolved elsewhere.


\section*{Acknowledgments}
We would like to thank D. V. Vassilevich for valuable comments on our previous version of the
manuscript, and Hwajin Um for
exciting discussions. Eune was supported by Basic Science Research Program through the
National Research Foundation of Korea(NRF) funded by the Ministry of
Education(2018R1D1A1B05050636).  Kim was supported by the National
Research Foundation of Korea(NRF) grant funded by the Korea
government(MSIP) (2017R1A2B2006159).


\bibliographystyle{JHEP}       

\bibliography{references}

\end{document}